\documentclass[twocolumn]{aastex61}
\pdfoutput=1 
\usepackage{amsmath,amstext}
\usepackage[T1]{fontenc}
\usepackage{apjfonts} 
\usepackage[figure,figure*]{hypcap}
\usepackage{natbib}

\shorttitle{AASTeX 6.1 Template}
\shortauthors{Tripathi et al.}

\begin{document}

\title{Origin and recovery from grand solar minima in a time delay dynamo model with magnetic noise as an additional poloidal source}

\author{Bindesh Tripathi}
\affiliation{Department of Physics, St. Xavier's College, Tribhuvan University, Kathmandu, Nepal}

\author{Dibyendu Nandy}
\affiliation{Department of Physical Sciences, Indian Institute of Science Education and Research, Kolkata, India}
\affiliation{Center of Excellence in Space Sciences India, IISER Kolkata, Mohanpur 741246, West Bengal, India}
\author{Soumitro Banerjee}
\affiliation{Department of Physical Sciences, Indian Institute of Science Education and Research, Kolkata, India}

\begin{abstract}
We explore a reduced Babcock-Leighton (BL) dynamo model based on delay differential equations using numerical bifurcation analysis. This model reveals hysteresis, seen in the recent mean-field dynamo model and the direct numerical simulations of turbulent dynamos. The BL model with 'magnetic noise' as an additional weak-source of the poloidal field recovers the solar cycle every time from grand minima, which BL source alone cannot do. The noise-incorporated model exhibits a bimodal distribution of toroidal field energy confirming two modes of solar activity. It also shows intermittency and reproduces phase space collapse, an experimental signature of the Maunder Minimum. The occurrence statistics of grand minima in our model agree reasonably well with the observed statistics in the reconstructed sunspot number. Finally, we demonstrate that the level of magnetic noise controls the duration of grand minima and even has a handle over its waiting period, suggesting a triggering effect of grand minima by the noise and thus shutting down the global dynamo. Therefore, we conclude that the 'magnetic noise' due to small-scale turbulent dynamo action (or other sources) plays a vital role even in Babcock-Leighton dynamo models.
\end{abstract}



\section{Introduction}

The number of sunspots on the solar surface waxes and wanes to produce a roughly 11-year solar cycle. These sunspots are highly magnetized regions, which are believed to be generated through magnetohydrodynamic (MHD) dynamos \citep{parker1955hydromagnetic, charbonneau2010dynamo}. The cyclic variation in the sunspot number is found to modulate several components of the space weather like total solar irradiance, coronal mass ejections, solar flares, solar winds, cosmic ray flux, and possibly the Earth's climate \citep{lean1995reconstruction,crowley2000causes,marsh2003solar}. Interestingly, the sunspots sometimes appear very few in number or do not appear at all for an extended period of time. Such periods of reduced magnetic activity is referred to as grand minima of which the Maunder Minimum was the last one observed from 1645 to 1715 AD \citep{eddy_maunder_1976}. It has been statistically justified that the Maunder Minimum is not an artifact of few observations but was observed 68\% of the days during this period \citep{hoyt_how_1996}. Several groups who have reconstructed the sunspot number, based on the studies of cosmogenic isotopes like ${Be}^{10}$ in ice cores and $C^{14}$ in tree rings suggest that such grand minima episodes had occurred in the past as well \citep{usoskin_grand_2007, steinhilber20129}. The reconstructed data shows that 20 grand minima had occurred in the last 9000 years, with the sun spending nearly 17\% of that time interval in such episodes \citep{usoskin2016solar}. The indirect proxies also suggest that even when no sunspots are observed, the magnetic cycle continues during grand minima, albeit at a subdued level \citep{beer1998active, fligge1999determination,miyahara2004cyclicity}. Such epochs have been recently confirmed as a special mode of the solar dynamo operation, distinct from the regular activity mode \citep{usoskin_evidence_2014}. We also highlight that the low magnetic activity periods like the Maunder Minimum phases have been realized in solar-type stars as well \citep{baliunas1995chromospheric} and their very existence has become an enigma for the solar and stellar physicists to understand them. The study of such minima is of utmost importance since it sheds light towards having a complete understanding of the working of the solar and the stellar dynamos and hopefully in predicting such events in the near future.\\

The global magnetic field of the sun can be divided into the toroidal field (in the azimuthal direction) and the poloidal field (in the meridional plane). The regeneration of one component of the magnetic field from the other one, mediated via plasma flows, is the idea of dynamo mechanism \citep{charbonneau2010dynamo,ossendrijver2003solar}. The poloidal fields are sheared due to the strongest differential rotation occurring at the base of the solar convection zone (SCZ) and produces the toroidal field. This process called $\Omega$-effect \citep{parker1955hydromagnetic} is further favored by a very low diffusivity at the base of the SCZ, which results in the amplification and the storage of the toroidal field. These toroidal flux tubes rise up to the solar surface due to magnetic buoyancy and pierce the surface at two regions called the bipolar active regions (also known as sunspot pairs), which are tilted with respect to the East-West direction. The decay of these tilted bipolar active regions, traditionally referred to as the Babcock-Leighton mechanism, is the only observed source for the regeneration of the poloidal field (to complete the loop of dynamo action) \citep{wang2009understanding, munoz2013solar, babcock1961topology} although mean-field alpha effect, small-scale turbulent dynamos, etc. are hypothesized to regenerate the poloidal field. These surface poloidal fluxes are transported to the tachocline region by the joint contribution of different flux transport mechanisms: meridional circulation, turbulent diffusion, and magnetic pumping. In Babcock-Leighton dynamos, the sources of the toroidal and the poloidal fields are spatially segregated as they operate at the base of the SCZ and near the solar-surface respectively. Such spatial segregation introduces time delays in the communication between two source layers and thus creates a memory even in the stochastic system. This memory has been used by solar physicists to predict the solar cycles \citep{yeates_exploring_2008, petrovay2010solar}.\\ 

The amplitudes of the solar cycles show a wide variability and their irregularities are modeled using mainly two approaches, viz. fluctuations in the poloidal field regeneration process and fluctuations in the flux transport time scale due to the magnetic back reaction on the fluid motion. These stochastic fluctuations result from the highly turbulent nature of the SCZ and hence are invoked in the solar dynamo models as fluctuations around a predefined mean value of some model's ingredients like the $\alpha$ effect \citep{hoyng_turbulent_1988, hoyng_effect_1994, choudhuri_stochastic_1992, charbonneau_intermittency_2004, usoskin_history_2013} or the meridional circulation \citep{charbonneau_stochastic_2000,lopes_solar_2009,karak_importance_2010}. Convective turbulence also introduces 'magnetic noise' that is directly associated with stochastic fluctuations in the mean electromotive force in mean-field dynamos\citep{brandenburg2008modeling}. These fluctuations cause an irregular change in the amplitude of the cycle, modulating it on decadal to centennial timescale although they are invoked at each correlation time shorter than the solar cycle timescale. These stochastically forced models are robust in the sense that they induce variations in the solar cycle over a wide range of parameters.\\

The other way of exploring the solar cycle variability that is recently being carried out is producing 3D magnetohydrodynamical simulations of the solar convection zone \citep{cossette2013cyclic,nelson2012magnetic,charbonneau2014solar}. Such simulations have become a great boon to understand the origins of grand minima and their subsequent recovery as they take into account the dual interaction between plasma flows and magnetic fields \citep{Augustson2015}. They have started exhibiting stable large-scale dynamo action and explaining detail underlying mechanisms of the dynamo \citep{ghizaru2010magnetic,brown2011magnetic,kapyla2012cyclic}. Another popular way of probing the global solar dynamo is to simplify and truncate its 3D MHD counterparts. Such truncated low-order dynamo models have been found to successfully mimic several observed solar variations \citep{mininni_simple_2001,pontieri_simple_2003,wilmot-smith_time_2006,passos2008low,cameroonschussler2017}. It also keeps physics transparent and offers faster and longer integration time as compared to its 2.5D and 3D counterparts, allowing us to study the long-term behavior of the solar activity. Here we employ a reduced model based on delay differential equations and stochastic fluctuations in the poloidal field source to study the long-term solar cycle variability. It is found that the usual fluctuations in the Babcock-Leighton dynamo parameters can push the sun into grand minima phases while the recovery from such phases requires some different poloidal field regeneration mechanism other than the Babcock-Leighton mechanism.\\ 

In the next section we briefly outline the time delay model. We present in section \ref{subsection hysteresis explanation} that the time delay Babcock-Leighton dynamo model, despite being a reduced model, captures physics so well that it successfully reproduces hysteresis phenomenon, which complicated mean-field models and 3D simulations of turbulent dynamos have revealed. We then present a possible mechanism (in section \ref{subsection magNoise}) for the recovery from such episodes and mimic several solar observations by considering the presence of a low-amplitude stochastic 'magnetic noise'. In the last section we conclude that the noise (due to small-scale turbulent dynamo action or other sources) can also shut down the global dynamo of the sun.


\section{Time delay dynamo model}

\begin{figure}[htb!]
\centering \includegraphics[width=0.4\textwidth]{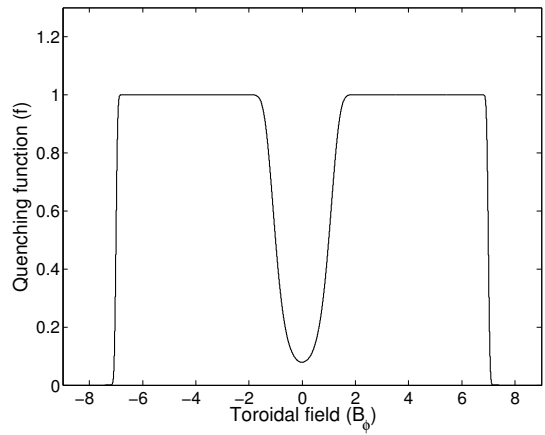}
\caption{\label{fig quenching} Profile of the Babcock-Leighton poloidal source quenching function $f$, with $B_{min}=1$, $B_{max}=7$ and equipartition field strength, $B_{eq}=1$ (all in arbitrary code units).}
\end{figure}

We adopt the delay model set up by \citet{wilmot-smith_time_2006}. The model was constructed by considering only the source and dissipative mechanisms in the dynamo equations. The kinematic mean-field dynamo equations were truncated and all space dependent terms were removed. Instead indispensable time delays in the communication between two segregated source layers in a Babcock-Leighton dynamo model were imbibed. The time delay dynamo equations are
\begin{eqnarray}
\frac{dB_\phi(t)}{dt}&=&\frac{\omega}{L}A(t-T_0)-\frac{B\phi(t)}{\tau} \label{toroidal time delay eqn} \\
\frac{dA(t)}{dt}&=&{\alpha_{BL} f(B_\phi(t-T_1))B_\phi(t-T_1)}-\frac{A(t)}{\tau} \label{Poloidal time delay eqn}
\end{eqnarray} \vspace{-0.7cm}
where the quenching factor $f$, nonlinearly approximated as\\

\begin{eqnarray}
\begin{aligned}
f=& \frac{[1+erf({B_\phi}^2(t-T_1)-{B_{min}^2})]}{2} \\ & \frac{[1-erf({B_\phi}^2(t-T_1)-{B_{max}^2})]}{2},
\end{aligned}
\end{eqnarray}
takes care of the quenching of the poloidal source when the toroidal field exceeds an upper threshold ($B_{max}$). This quenching results from the Lorentz feedback of the strong toroidal fields on helical turbulence in the mean field dynamos while it results from the ineffective Coriolis force on the magnetically buoyant toroidal flux tubes which forms bipolar sunspots, in the Babcock-Leighton dynamo, without significant tilts \citep{dsilva_theoretical_1993,fan_origin_1993}. 
Here, the Babcock-Leighton poloidal source has a lower operating threshold ($B_{min}$) as well, below which the hydrodynamical forces dominate over the magnetic buoyancy, resulting in no formation of sunspots and consequently no contribution to the poloidal source. The profile of quenching function $f$ is shown in Figure \ref{fig quenching}.

The time delays $T_0$ and $T_1$ are the time required for the process of conversion of the poloidal field to the toroidal field and vice-versa. In Equation (\ref{Poloidal time delay eqn}), $\alpha_{BL}$ is directly related to dynamo number and represents the tilt angle of bipolar active regions. For a constant value of $\alpha_{BL}$, we realize a strictly periodic solution. Being motivated by the observational fact that the tilt angles of the bipolar active regions are scattered around a mean value given by Joy's law distribution, we use stochastic fluctuations in $\alpha_{BL}$ \citep{howard_axial_1991,dasi-espuig_sunspot_2010}. The physical reason behind this dispersion is that while strong toroidal flux tubes are rising up due to magnetic buoyancy, they are randomly buffeted in the turbulent SCZ, which inherently imparts a random component to the systematic tilt angle. The fluctuations in $\alpha_{BL}$ are introduced at each coherence time $\tau_{cor}$ (\citet{hazra_stochastically_2014}) as
\begin{eqnarray} \label{stochastic_alpha}
\alpha_{BL} = \alpha_{mean} \left[1+\frac{\delta}{100} \sigma (t,\tau_{cor}) \right]
\end{eqnarray}
where $\alpha_{mean}$ stands for a value of $\alpha_{BL}$  around which stochasticity is forced at $\delta$ percentile level. This $\alpha_{mean}$ is adjusted using numerical bifurcation analysis (presented in the next section) in order to reproduce grand minima episodes interspersed between regular activity cycles with varying amplitudes. We use $\sigma (t, \tau_{cor})$ as a uniform random function \citep{hazra_stochastically_2014} lying within the interval [-1, +1] and changing its value at each correlation time $\tau_{cor}$. We numerically integrate stochastic delay differential equations \ref{toroidal time delay eqn} and \ref{Poloidal time delay eqn} by amalgamating MATLAB's dde23 package and random number generating function.

\begin{figure*}[htb!] \label{fake recovery and unsuccessful} 
\centering \includegraphics[width=0.95\textwidth,height=0.3\textwidth]{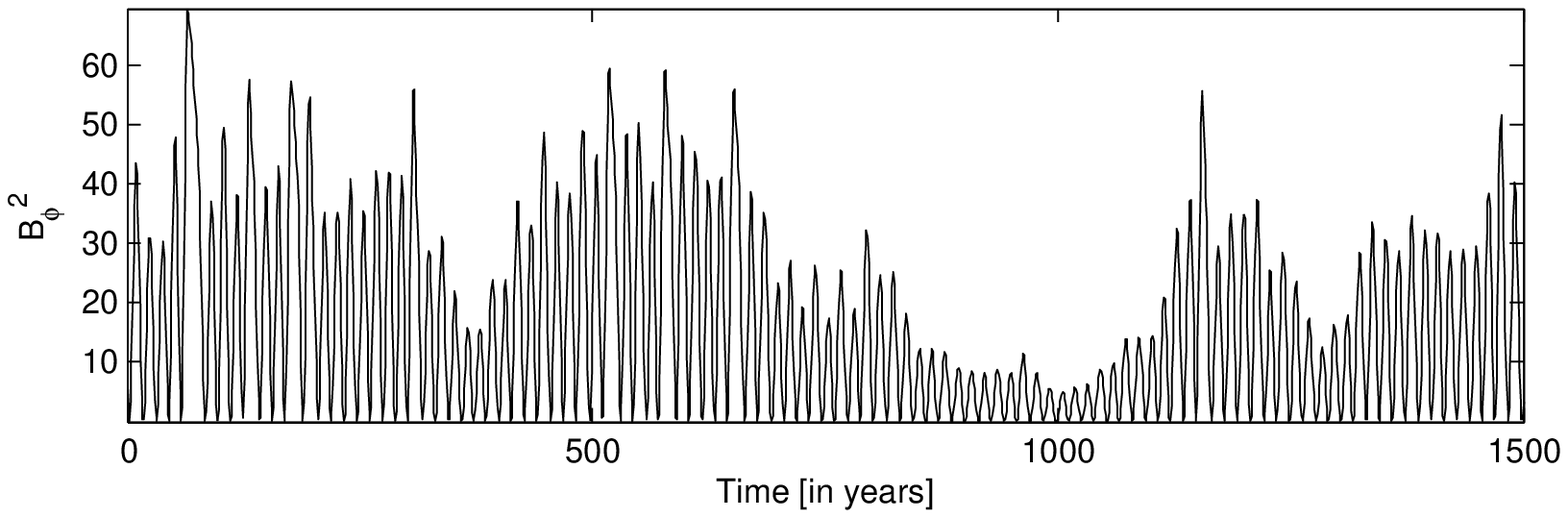}\vspace{0.4cm}
\includegraphics[width=0.95\textwidth,height=0.33\textwidth]{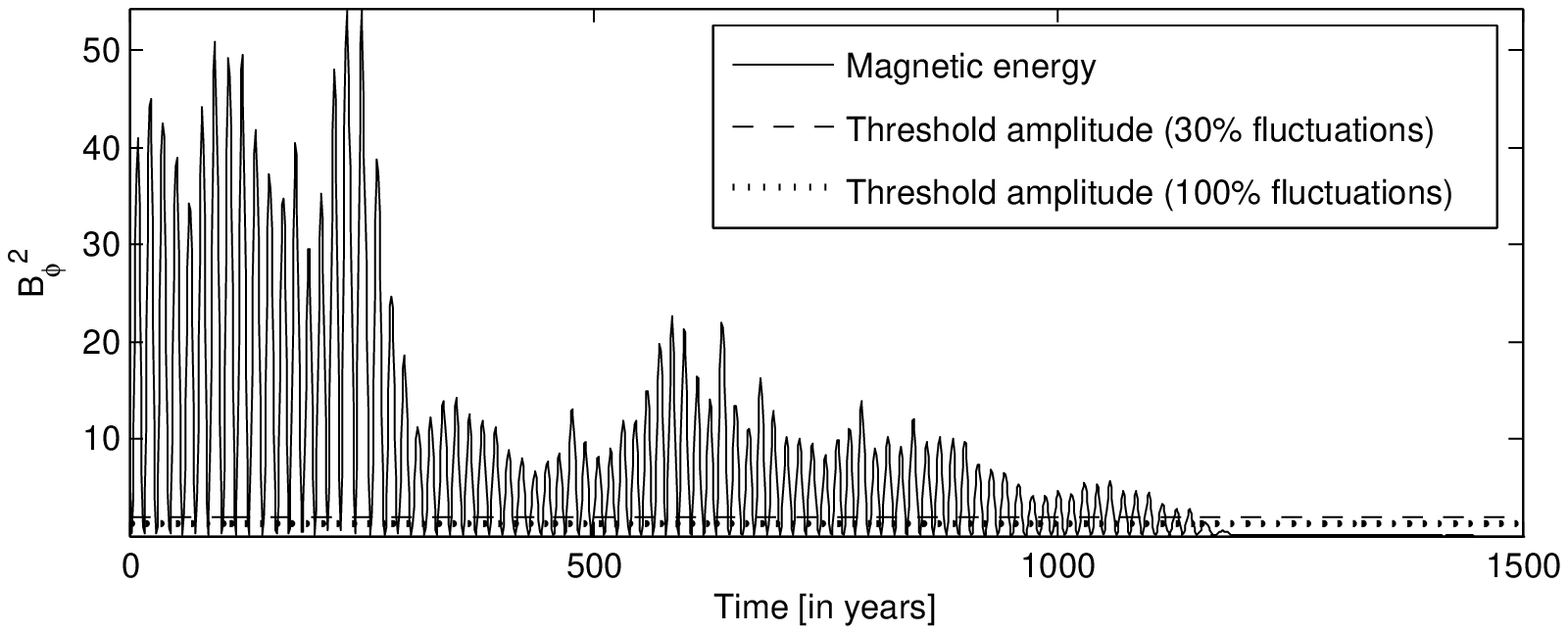}  
\caption{(a) Time series of toroidal magnetic energy (a proxy of sunspots) with an upper operating threshold (i.e. $B_{max}=7$) but without a lower operating threshold (i.e. $B_{min}=0$); (b) Same as above but with a non-zero lower operating threshold ($B_{min}=1$). All other parameters are set at $\tau=15$, $T_0=2$, $T_1=0.5$, $\omega/L=-0.34$, $B_{max}=7$ and $\alpha_{mean}=0.17$, i.e. ${N_D}^{mean}=13$. In the latter case, the sunspot cycle's amplitude once falls below a certain threshold can never recover from it. The threshold line for 100\% fluctuations in $\alpha_{BL}$ (or $N_D$) is slightly lower as compared to the threshold line for its 30\% fluctuations; see in the text to see how these thresholds, which need not necessarily be the threshold for sunspot formation, are quantified.} 
\end{figure*}

The ratio of the product of the source terms for toroidal and poloidal fields generation to the respective diffusion terms in the dynamo equations is commonly termed as dynamo number. It is a measure of the efficiency of the dynamo mechanism and is given as
$N_D = \omega \alpha_{BL} \tau ^2/L$. The expected diffusion timescale ($\tau = L^2/\eta$) is 13.8 years, using typical values of diffusivity ($\eta = 10^{12}$ $cm^2s^{-1}$) and the length of the SCZ ($L = 0.3R_0$, $R_0$ is the solar radius). We refer to \citet{hazra_stochastically_2014} and \citet{wilmot-smith_time_2006} for a detailed parametric consideration. In brief, the ratio of $B_{max}/B_{min}$ is taken to be 7 as the upper threshold $B_{max}$ above which the active regions appear without significant tilts is on the order of $10^5$ G and the lower operating threshold $B_{min}$ below which hydrodynamical forces in the SCZ dominate over the magnetic buoyancy and subsequently no formation of sunspots take place is on the order of $10^4$ G. The time delay for the generation of the toroidal field from the poloidal field (by considering magnetic pumping as an effective flux transport mechanism) is taken as 2 years and the magnetically buoyant flux tube rising timescale is considered 6 months. Initial conditions for $A$ and $B_\phi$ to solve the stochastic delay differential equations are taken to be $(B_{min}+B_{max})/2$. We now present the features of a stochastically forced time delay Babcock-Leighton dynamo model and proceed to explore the possible mechanisms for the onset and recovery of solar cycle from grand minima.


\section{Results and discussions}

\begin{figure*}[htb!] \label{Bifurcation diagrams} 
\centering \hspace{0.7cm} \includegraphics[width=0.75\textwidth]{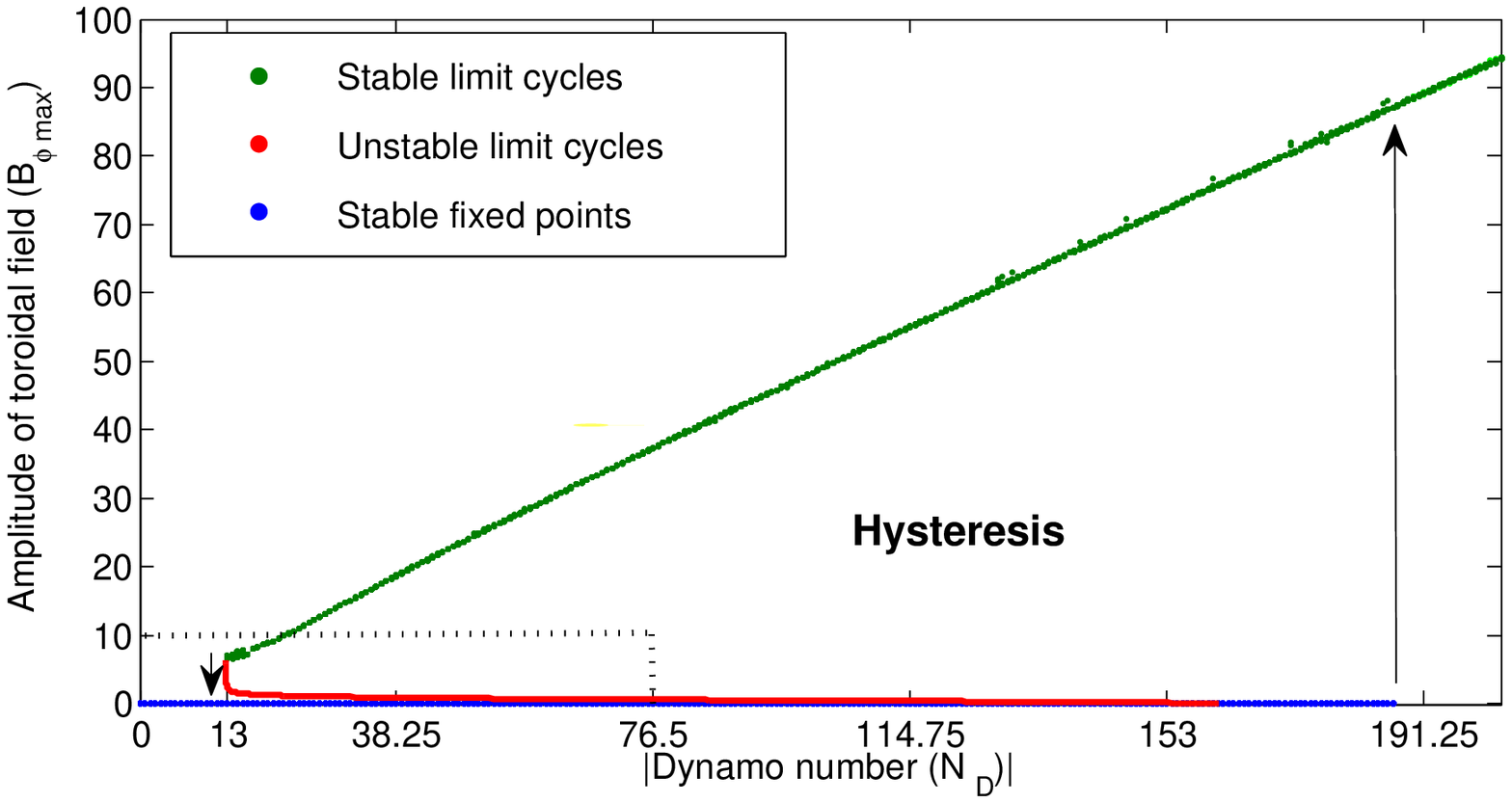} \includegraphics[width=0.7\textwidth]{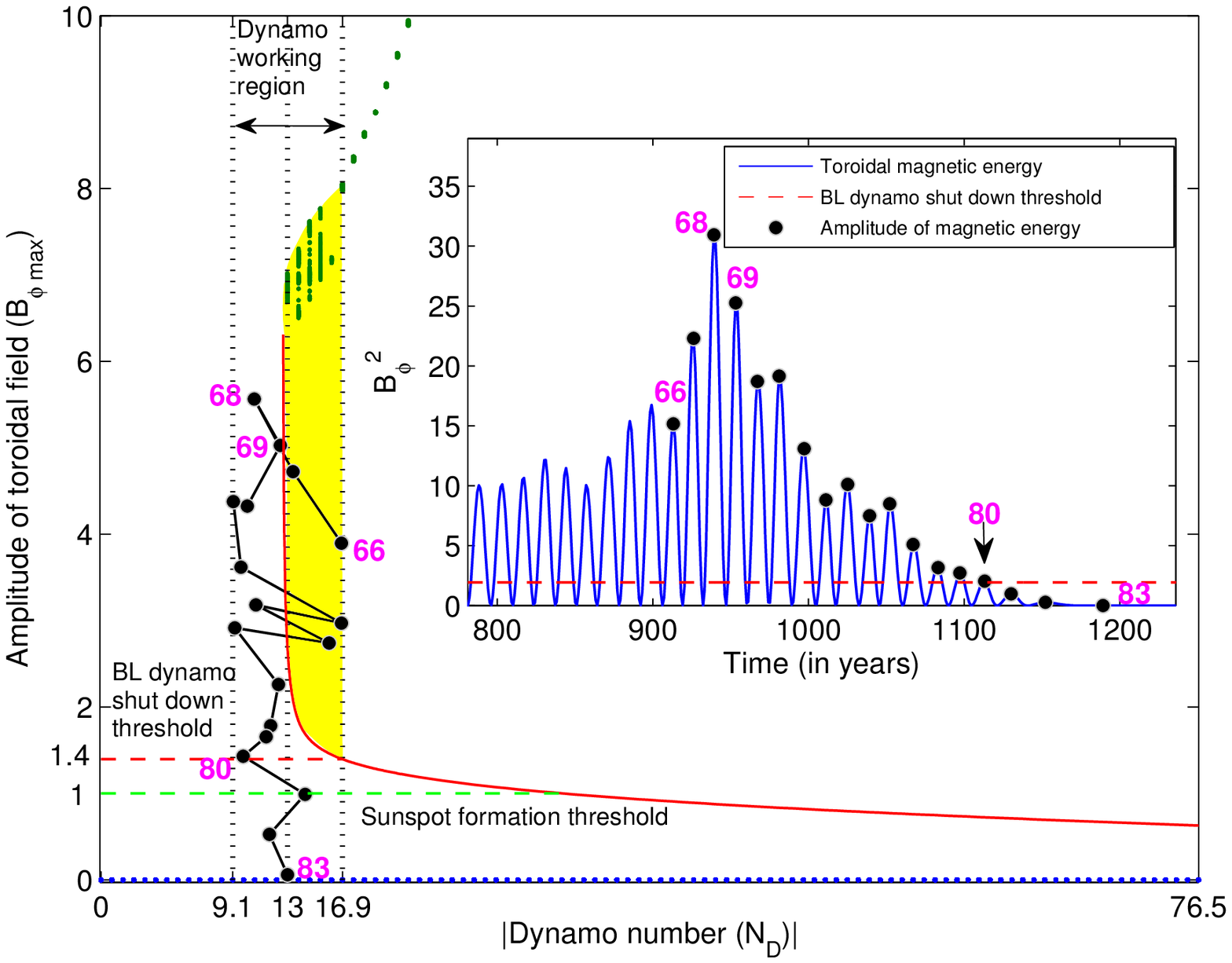}
\caption{(Upper panel) Bifurcation diagram of the amplitude of toroidal field in simulated cycles. The \textit{red curve} (unstable limit cycle) separates the basin of attraction of the oscillating solutions (stable limit cycles; shown in \textit{green dots}) and the decaying solutions (stable fixed points; shown in \textit{blue dots}). Hysteresis can be vividly observed as bistable solutions exist for the parameter range, $N_D \in [12.6, 183.6]$. The dotted rectangular region is zoomed and shown in the lower panel. (Lower panel) The region within the \textit{dotted} vertical lines at $N_D=9.2$ and $N_D=16.9$ is the dynamo working region when 30\% fluctuation is used at $\alpha_{mean}=0.17$ (i.e. ${N_D}^{mean}=13$). The \textit{dashed} red horizontal line is the threshold below which if the amplitude of the cycle falls, the cycle never recovers if the Babcock-Leighton (BL) process alone is considered as poloidal source. The \textit{dashed} green horizontal line, which is the threshold for sunspot formation, lies below the BL dynamo shut down threshold. The \textit{yellow shaded} region represents the basin of attraction for the oscillatory solutions. The numbers in pink are the corresponding cycle's number counted from the beginning of the simulation. The plot shows the typical onset of grand minimum due to the successive weakening of the poloidal source via fluctuations in BL $\alpha$-effect. The corresponding time evolution of magnetic energy for the fluctuations in dynamo number (noted at cycle maxima) with its effect on the amplitude of toroidal field (shown by black filled circles) is illustrated on the right side. The cycle cannot recover from the grand minimum phase once it sets on.}
\end{figure*}

Several Babcock-Leighton (BL) dynamo models have explored grand minima without considering the lower operating threshold (i.e. $B_{min} = 0$). Although such consideration seems to explain the recovery of the solar cycle from grand minima (see Figure \ref{fake recovery and unsuccessful}, upper panel), they are unphysical and unrealistic. The reason behind this is that the very weak toroidal flux tubes would not be able to find their path to the solar surface due to the dominating effect of the hydrodynamical forces over the magnetic buoyancy on such weak flux tubes and hence these flux tubes do not form sunspots. The Babcock-Leighton source is then expected to switch off, giving rise to a catastrophic decay of the solar cycle unless there is some other mechanism that operates to restart the global dynamo. Also the incorporation of lower operating threshold due to magnetic buoyancy is found to limit the cycle amplitude in Babcock-Leighton dynamo models and thus is necessary to consider in all such kind of dynamo models \citep{nandy_constraints_2002}. \citet{hazra_stochastically_2014} showed that when we introduce the lower operating threshold in the model, the cycle never recovers from a grand minimum once the amplitude of the cycle of the toroidal field falls below a threshold. This is shown in Figure \ref{fake recovery and unsuccessful} (lower panel). To quantify the dynamo shut down threshold and to explore the physics, in great detail, behind such unsuccessful recovery, we draw a bifurcation diagram (a plot of a variable versus a parameter) numerically. Figure \ref{Bifurcation diagrams} (upper panel) is the plot of the amplitude of the toroidal field of a stable cycle for different values of the $N_D$ parameter. The green and blue dots in the figure represent oscillating and decaying solutions respectively. The fluctuations in $N_D$ (or correspondingly $\alpha_{BL}$) causes the sun to transit across the red curve that separates the basins of attraction of the oscillating solutions and the decaying solutions. This stochastic transition causes the solar cycle amplitude to grow and decay irregularly from cycle to cycle. So, the two kinds of solutions: growing and decaying solutions, are irregularly realized depending on the prehistory of the variation of $N_D$.\\

The bifurcation diagram is drawn integrating the set of delay differentials equations initially with a constant very low value of $N_D$ for a sufficiently long time to get a constant amplitude of cycles. (It should be kept in mind that all cycles decay for very low values of $N_D$.) Afterwards, the equations are integrated repeatedly with increasing values of $N_D$ , along with feeding the final conditions of previous integration as initial conditions for the next integration. After removing the initial transients in each integration, we note the steady amplitudes of the cycles. This process is continued till we obtain oscillatory solutions. Afterwards we follow the same procedure but with decreasing values of $N_D$ till we observe decaying solutions. It can be clearly seen in Figure \ref{Bifurcation diagrams} (upper panel) that the time delay dynamo model exhibits hysteresis as the decaying solutions change into oscillating ones for all initial conditions at $N_D=183.6$ while the oscillating solutions reverses into decaying ones for all initial conditions only at $N_D=12.6$. \\


\subsection{ \label{subsection hysteresis explanation} Hysteresis explains grand minima and regular activity as distinct modes of dynamo operation} 

Reconstructed solar activity record reveals that the sun operates in two distinct modes of activity: the main regular activity mode and the grand minima mode when the sun remains quite. \citet{usoskin_evidence_2014} showed, at a high level of confidence, that the grand minima mode cannot be explained by considering a low-activity tail (random fluctuations) of a single regular solar-activity mode. These two distinct modes can be directly related to hysteresis phenomenon in a dynamo model where bistable solutions are possible for a certain parameter regime and the sun is pushed irregularly towards either of the solution's basin of attraction depending on the value of dynamo parameter. The system of delay differential equations of dynamo model that we have considered here shows hysteresis. Similar behavior was found to exist by \citet{kitchatinov_dynamo_2010} using a nonlinear mean-field dynamo model with flux transport coefficients depending on the strength of magnetic field. Recently, \citet{karak_hysteresis_2015} found hysteresis phenomenon in 3D numerical simulations of completely different dynamo set up- the turbulent dynamos, with the irregular intermittency of magnetic cycles between relatively high amplitude oscillations and the low activity epochs. However, the sluggish 3D computations hindered them from analyzing the distribution function of a huge number of magnetic cycles observed in the simulations. We present such analysis, utilizing our time delay model, in the section \ref{subsection magNoise}.\\

Now we move on to adjusting the value of $\alpha_{mean}$ (or equivalently the mean dynamo number, ${N_D}^{mean}$) of Equation (\ref{stochastic_alpha}). The high value of ${N_D}^{mean}$ is not feasible for the presently existing sun as suppressed activity periods (grand minima) are hardly realized with such parametric consideration. The reason behind this is that a high value of ${N_D}^{mean}$ offers an extremely low possibility for the cycle's amplitude to lie below the red separating curve and thus low-activity epochs (grand minima) are rarely realized as $N_D$ fluctuates. Physically explaining, the high value of ${N_D}^{mean}$ means that the dynamo is hugely efficient and hence, the toroidal field's amplitudes are also mostly very high. Such scenario could have persisted in the distant past when the solar dynamo was highly efficient due to the faster rotation rate of the sun \citep{reiners2012observations}. (Note that it is often believed, the faster rotating stars have a stronger differential rotation, making the dynamo more efficient.) With the evolution of time, the loss of its angular momentum through a magnetically coupled solar wind \citep{kraft_studies_1967,hartmann_rotation_1987} might have slowed it down, bringing it to the left end of the bifurcation diagram, near the dynamo onset region where they exhibit intermittent grand minima episodes \citep{metcalfe2016stellar,van2016weakened}. This picture is justifiable as the weak magnetic activities have actually been observed only in old stars \citep{wright_we_2004}. On the other hand, the very low value of ${N_D}^{mean}$, however, traps the sun often into the basin of attraction of the decaying solutions (i.e. on the left and below the red curve in Figure \ref{Bifurcation diagrams}, upper panel), producing a lesser number of growing solar cycles with mostly decaying cycles (grand minima epochs) and hence the solution is again non-solar like. However, we speculate that the sun might exhibit such behavior as its rotation rate further slows down in the distant future. So, for the present case of the sun, we choose a sub-critical value of $\alpha_{mean}$ (=0.17; ${N_D}^{mean}=13$ shown by a middle dotted-vertical line that nearly overlaps with the vertical section of the red separating curve; see Figure \ref{Bifurcation diagrams}, lower panel) in such a fashion that the size of the basins of attraction for the decaying and growing amplitudes' cases are nearly equal. Such sub-critical choice of dynamo number is also supported by the recent stellar evidences that hint that the sun may be in transition phase from a fast-rotating and magnetically active type star to a slow-rotating and magnetically weak type star \citep{metcalfe2016stellar, van2016weakened}. \\

\begin{figure*}[htb!]
\centering \includegraphics[width=0.7\textwidth]{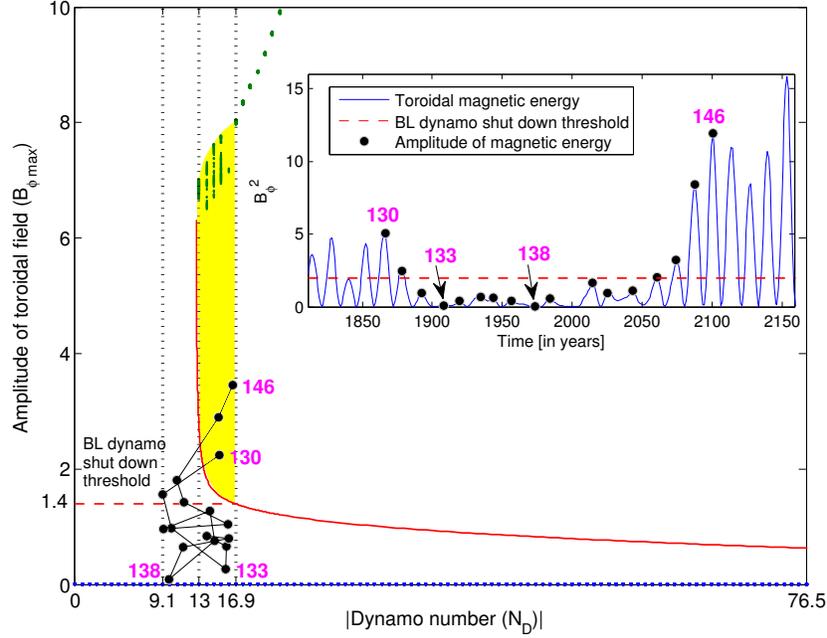}
\caption{\label{successful recovery} The same bifurcation diagram as in Figure \ref{Bifurcation diagrams} (lower panel) with fluctuating values of $N_D$ noted at the cycle maxima of the time series (shown on the right). Here the cycle recovers every time it hits grand minimum because of the consideration of magnetic noise in the Babcock-Leighton dynamo model as a weak-field poloidal source. The \textit{dashed} horizontal line is the threshold below which if the amplitude of the cycle falls, the Babcock-Leighton poloidal source alone cannot recover the cycle. The numbers in \textit{bold} are the corresponding cycle's number when the amplitudes are noted. The plot shows the typical onset and exit of grand minimum due to the additive magnetic noise. The parameters are set at $\alpha_{mean}=0.17$ (i.e. ${N_D}^{mean}=13$), level of fluctuation in $\alpha_{BL}$ ($\delta=30\%$) and the level of magnetic noise invoked is about 3\% of the average of the toroidal field's amplitude during regular activity cycles.}
\end{figure*}

The BL dynamo shut down threshold in Figure \ref{Bifurcation diagrams} (lower panel) was drawn with the threshold ${B_{\phi}}_{max}$ corresponding to the value of ${B_{\phi}}_{max}$ on the red separating curve where it is intersected by the rightmost dotted-vertical line ($N_D=16.9$; it is the extreme most fluctuated value of $N_D$ under its 30\% fluctuation). Even such most effective BL source cannot push the cycle back to the (yellow) basin of attraction for growing solutions once the amplitude of the cycle falls below the threshold and thus the cycle never recovers. We observe that this threshold can be lowered (or pushed up) by considering slower (or faster) magnetic pumping speed. In case it is pushed up, the cycles are more prone to show low activity epochs. The rationale behind this is that, as we are in advection-dominated regime ($\tau>>T_0+T_1$), the faster pumping speed implies the lesser time for the shearing of the poloidal field, which ultimately produces weaker toroidal field. Even when the BL source is solicited to act most effectively (i.e. $N_D=16.9$), this weaker toroidal field, in turn, regenerates poloidal field weaker than what we started with if the toroidal field's amplitude is lower than the dynamo shut down threshold. This implies that we should start the cycle with a stronger poloidal field if we want the dynamo to keep on working in the case of increased pumping speed. This immediately sets a higher threshold for polar field to keep the dynamo operational, which obviously gets engraved in the amplitude of the next cycle's toroidal field. Thus we justify that the increased pumping speed in advection-dominated regime pushes the dynamo shut down threshold upward. So, a change in flux transport timescale can also be a culprit for triggering grand minima.\\

Figure \ref{Bifurcation diagrams} (lower panel) shows the variations of the amplitude of the toroidal field of different cycles, in a single realization, with the fluctuating $N_D$ parameter, noted when the toroidal field cycle gains its peak value.  This shows that the fluctuations in $N_D$ can trigger grand minima episodes but its fluctuations alone cannot recover the cycle.  The unsuccessful recovery of the cycle, with the Babcock-Leighton poloidal source alone, can be interpreted physically as the weakening of the poloidal field source for an extended period of time (i.e. $\alpha_{BL}$ taking lower value than the $\alpha_{mean}$) causing the toroidal field amplitude to decline below a threshold after which the diffusive mechanism dominates over the poloidal flux regeneration mechanism through Babcock-Leighton process. Thus the solar cycle amplitude keeps on falling and finally goes below the sunspot formation threshold suffering a catastrophe. It implies that an additional poloidal source is essential for the successful recovery of the cycle. We, therefore, consider 'magnetic noise' as a possible weak source of poloidal field. Under the addition of such weak source, we aim to find a successful recovery of the cycle and also persisting hysteresis phenomenon that existed earlier in the time delay dynamo model with Babcock-Leighton as the only poloidal source.\\

\begin{figure*}[htb!]
\centering \includegraphics[width=0.85\textwidth]{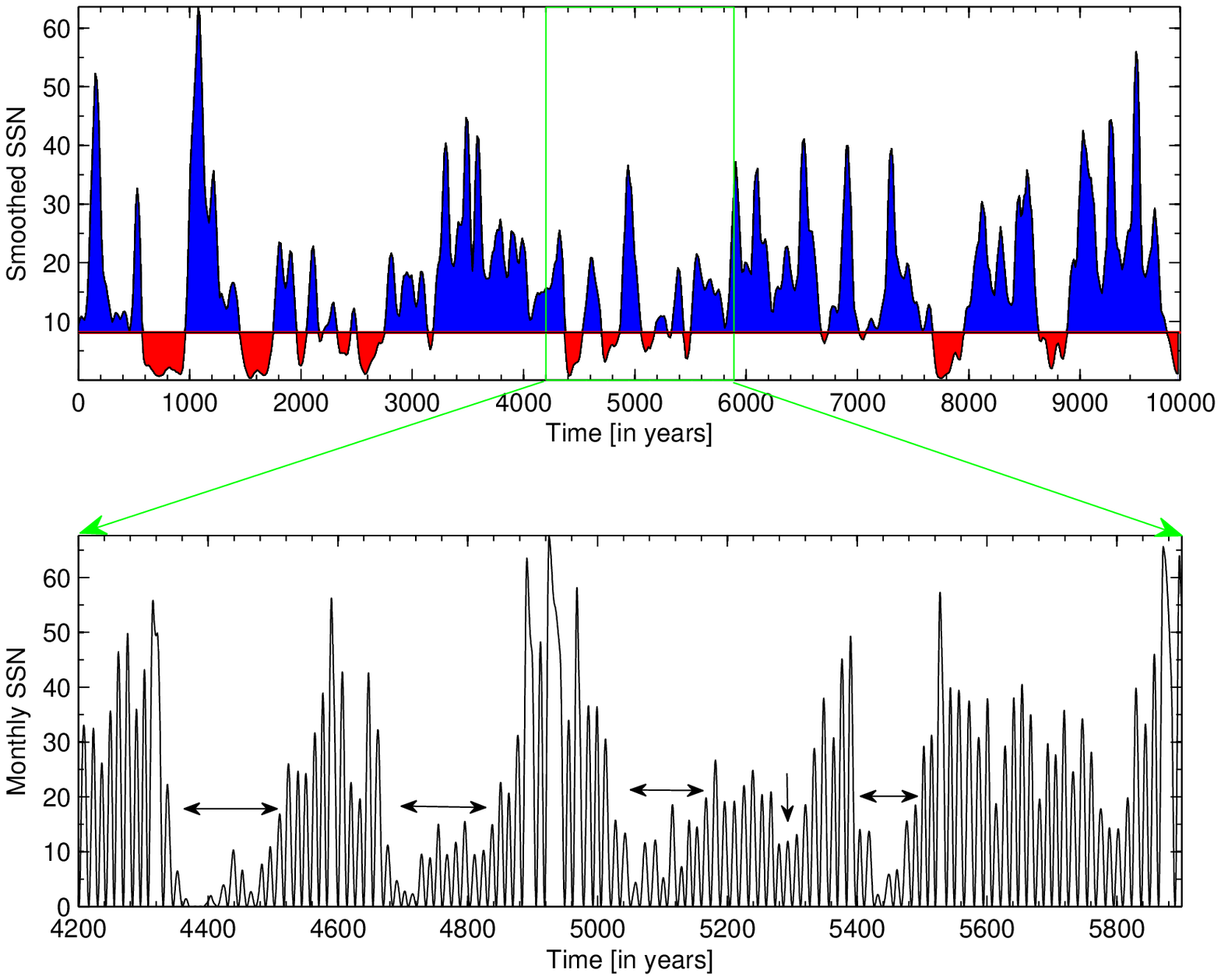}
\caption{\label{smoothSSN} (Upper panel) Temporal variation of the smoothed SSN from 10,000-year simulation. Red shaded regions below the horizontal line represent grand minima episodes. (Lower panel) Variation of monthly SSN for a selected duration of 1800 years. The arrows show the grand minima epochs. The level of additive noise imposed here is about 5\% of the average of toroidal field's amplitude during regular activity cycles. The parameters are set at $\omega/L=0.34$, $\alpha_{mean}=0.17$ (i.e. ${N_D}^{mean}=13$). Note the shortest grand minima, captured by the Gleissberg filter, that occurred at around time = 5300 years.}
\end{figure*}


\subsection{ \label{subsection magNoise} 'Magnetic noise' as a possible mechanism for the recovery from grand minima} 

'Magnetic noise' indefinitely abounds in the solar convection zone, especially near the sub-surface layers of the sun where the convection is much more turbulent. The small-scale dynamo action arising due to the extremely turbulent convection zone \citep{meneguzzi1989turbulent,nordlund1992dynamo} produces a magnetic field, which is highly inconsistent in space and time. Such noise pertains to the fluctuations in the mean electromotive force of the dynamo \citep{brandenburg2008modeling} and continuously regenerates and replenishes the magnetic field in the upper layers of the sun \citep{cattaneo_interaction_2003,hagenaar2003properties,charbonneau_intermittency_2004}. This field despite being very weak can be enough to feed the dynamo for its survival during the grand minima and eventually can provide enough strength to the dynamo for climbing back to the regular activity cycles. Although the magnetic noise preferably exists due to turbulent convection, we do not restrict ourselves to it for the generation of noise and thus hint that it might accrue due to some other sources as well.\\

\citet{charbonneau_multiperiodicity_2001}, in the case of a one-dimensional iterative map, has used time delay feedback mechanism inherent in the Babcock-Leighton dynamo along with a low-amplitude stochastic noise as an additional source of the poloidal field. The map considered there has a finite basin of attraction for oscillatory solutions. In the bifurcation diagram (see Figure \ref{Bifurcation diagrams}, lower panel), the time delay model with poloidal source as Babcock-Leighton type alone also shows similar finite basin of attraction. (It is because the finite percentile fluctuations in $N_D$ chops the bifurcation diagram in such manner that, within our dynamo working region from $N_D = 9.1$ to $N_D = 16.9$, there exists a finite basin of attraction for growing solutions.) However, our model presents a Hopf bifurcation instead of a period doubling bifurcation, seen in the one-dimensional iterative map. The stochastic fluctuations in $N_D$ bring the system sometimes very close to the boundary of the basin of attraction of oscillating solutions, after which further fluctuations in the $N_D$ parameter can nudge the system out of the basin, showing the quiescent phases afterward. Once the quiescent phase sets on, the fluctuations in $N_D$ alone cannot bring the cycle back to regular activity cycles. So, we test the recovery of the cycle after incorporating an additive 'magnetic noise' of low amplitude, which is stochastic in nature.\\

Let us recall that the Babcock-Leighton source term already has an unavoidable multiplicative noise due to the effect of convective turbulence on the rising toroidal flux tubes \citep{longcope2002orientational}. This stochasticity primarily governs the amplitude fluctuation of each cycle while the additive noise, resulting from small-scale turbulent dynamo action, provides a seed field for the dynamo. The Equation (\ref{Poloidal time delay eqn}) then becomes
\begin{eqnarray}
\frac{dA(t)}{dt}={\alpha_{BL} f(B_\phi(t-T_1))B_\phi(t-T_1)}-\frac{A(t)}{\tau}+\epsilon(t)
\end{eqnarray}
where $\epsilon(t)$ is uniform white noise in time \citep{charbonneau_multiperiodicity_2001,charbonneau_intermittency_2004,charbonneau2007fluctuations} and has zero mean. Since the noise has zero mean, it sometimes builds up the poloidal field (when the noise and the poloidal field have the same sign) and at other times, reduces the existing dipolar field of the sun (when the noise and the poloidal field have the opposite sign). We found that the incorporation of such magnetic noise in the model brings the cycle back to oscillation (see Figure \ref{successful recovery}) after some dynamo growth time, which depends on the level of noise imposed. It is observed that the amplitude of the field changes its polarity even during the quiescent phases although far less regularly than during the active phases, suggesting that even during the quiescent phases, the stochastically injected poloidal fields due to small-scale dynamo action (or other sources) are still being brought down at the tachocline region where they are sheared due to the differential rotation to regenerate the toroidal field. Thus even after the switching off of the Babcock-Leighton source, the magnetic cycle continues. As the stochastically injected poloidal fields are continuously being sheared at the base of SCZ, the amplitude of toroidal field, after certain number of cycles, build up and become strong enough to create sunspots and thus the sun is pulled back from a grand minimum.\\

A time series of smoothed cycle-averaged sunspot number from an additive-noise incorporated Babcock-Leighton dynamo model is shown in Figure \ref{smoothSSN} (upper panel). The smoothing has been done using the same procedure used by \citet{usoskin2016solar}, i.e. we first average the monthly SSN of each cycle over the cycle period and apply the Gleissberg low pass 1-2-2-2-1 filter. The red shaded regions below the horizontal line signify grand minima, which are defined as the events during which the smoothed SSN falls below 50\% of the average value of smoothed  SSN for at least 3 consecutive decades (Note that this is the revised definition of grand minima adopted by \citet{usoskin2016solar}). The lower panel of Figure \ref{smoothSSN} portrays the temporal variation of original SSN. The occurrence statistics of grand minima in our model agrees reasonably well with its statistics in the reconstructed data. We observe 16 grand minima in Figure \ref{smoothSSN} in a typical 9000-year simulation whereas the reconstructed data shows 20 grand minima in the last 9 millennia \citep{usoskin2016solar}. However, we note that the occurrence statistics of grand minima can easily increase or decrease depending upon the tuning of model's parameter.\\

\begin{figure*}[htb!]
\centering\includegraphics[width=0.6\textwidth]{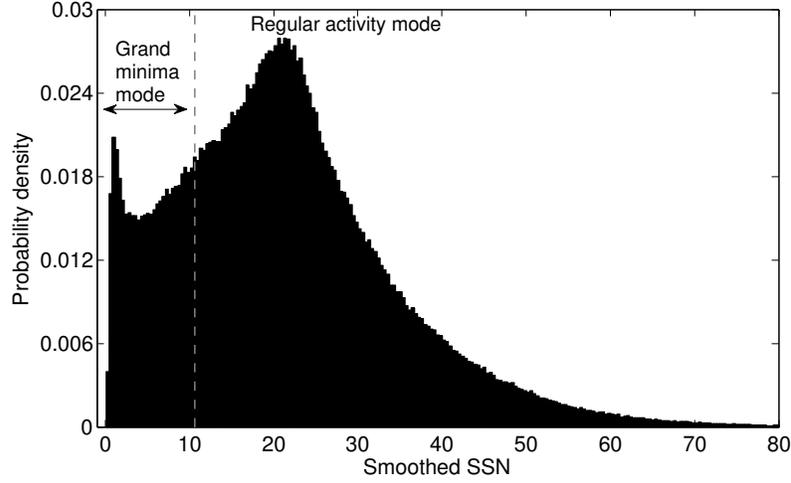}
\caption{\label{Bimodal_distribution} Bimodal distribution of probability density of magnetic energy (a proxy of sunspot number) shown by the time delay Babcock-Leighton model with magnetic noise as an additional weak poloidal source. The peak on the left suggests that the grand minima mode is a distinct mode of solar activity, apart from the regular activity mode (represented by the larger peak). This bimodal distribution is clearly due to the persistence of hysteresis in the model. The simulation was run for 10 million years by setting the parameters at $\alpha_{mean}=0.174$ (i.e. ${N_D}^{mean}=13.3$), level of fluctuations in $\alpha_{BL}$ was $30\%$ ($\delta=30\%$) and the level of additive noise is about $5\%$ of the average of toroidal field's amplitude during regular activity cycles.}
\end{figure*}

\begin{figure*}[htb!]
\includegraphics[width=0.44\textwidth]{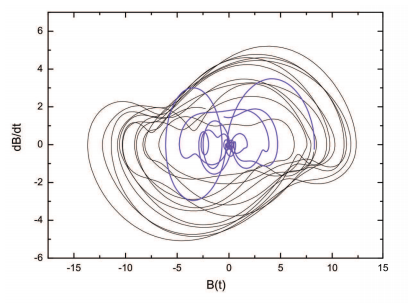}\includegraphics[width=0.5\textwidth]{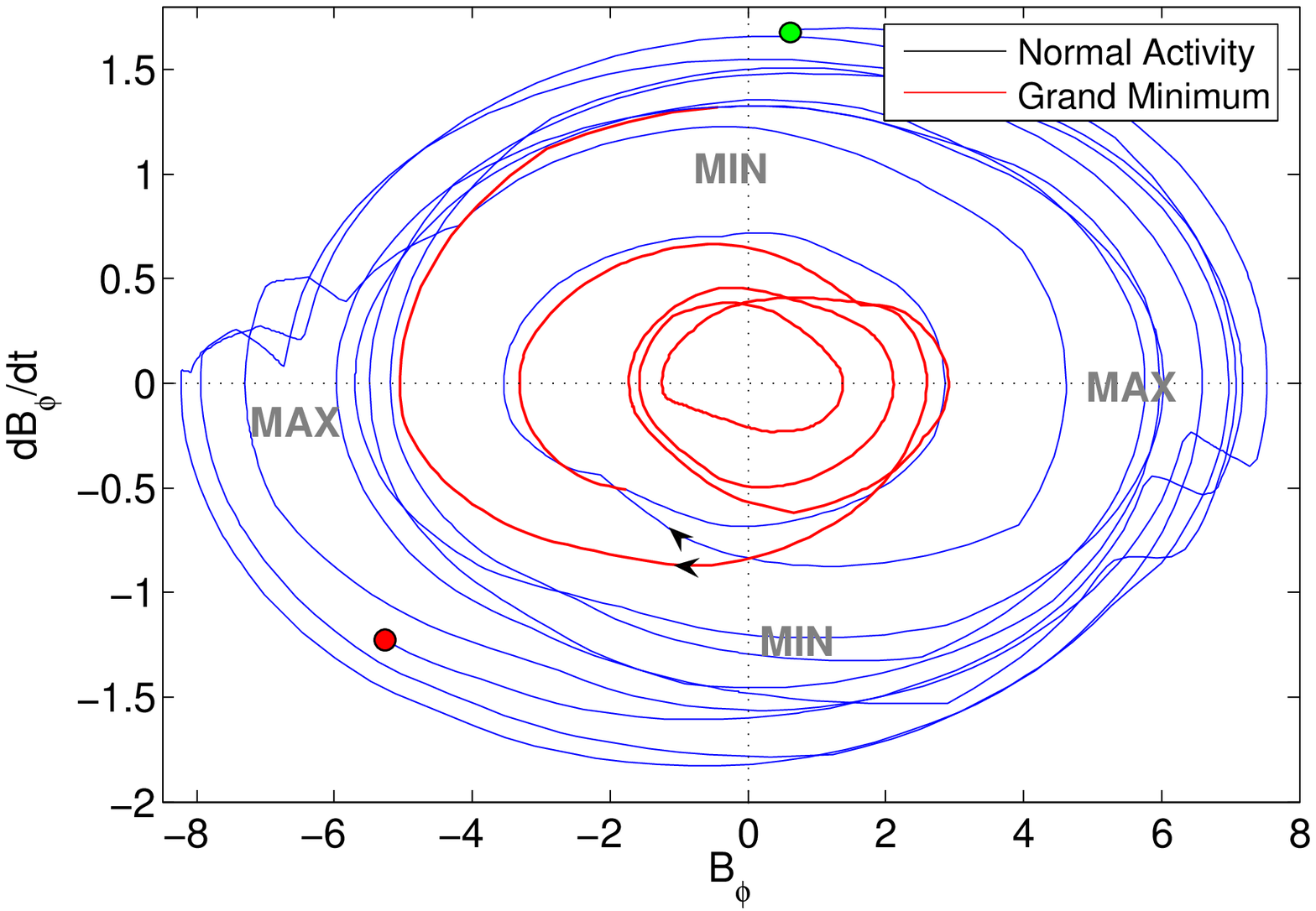} 
\centering \includegraphics[scale =0.75]{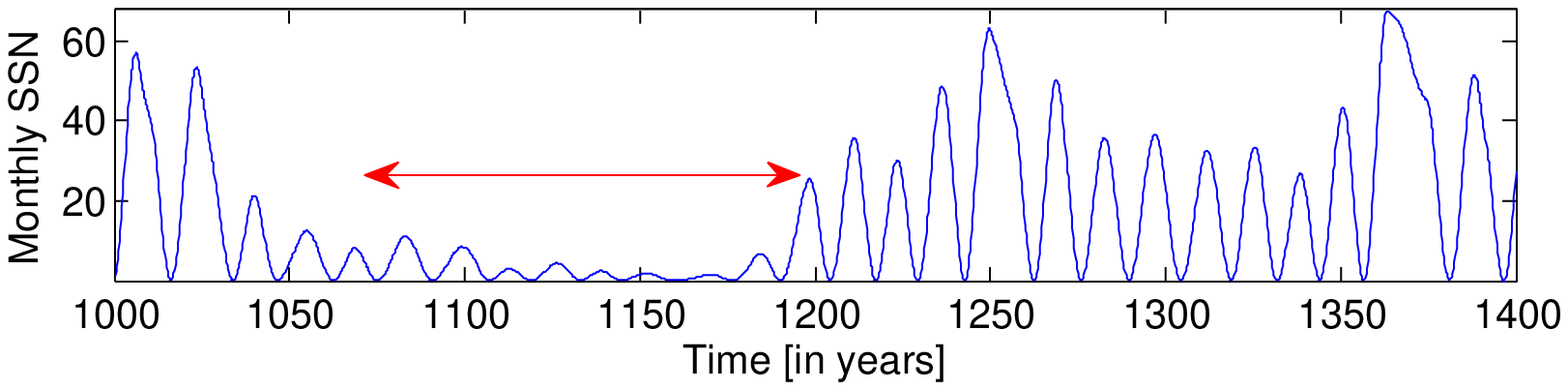}
\caption{\label{Phase space collapse} (Upper panel; on the left) Phase space of toroidal magnetic field built by
\citet{lopes_oscillator_2014} utilizing the directly observed sunspot number for the last 400 years. The curve in blue shows a collapse in the phase space, realized during the Maunder minimum period; (Upper panel; on the right) Phase space collapse seen in the simulation of a typical 400 years when the Sun went through a grand minimum phase. The green and red circles represent the beginning and the end of the considered 400 years. The cycles with large amplitudes in our model rise faster while their declining phases are lethargic. The simulation was performed using the parameters: $\alpha_{mean}=0.17$ (${N_D}^{mean}=13$), level of fluctuations in $\alpha_{BL}$ was $30\%$ and the level of additive noise invoked was about $3\%$ of the average of toroidal field's amplitude during regular activity cycles. (Lower panel) The corresponding temporal variation of SSN for the selected 400-year simulation in the upper panel.}
\end{figure*}

\begin{figure*}[htb!]
\includegraphics[width=0.48\textwidth, height=0.385\textwidth]{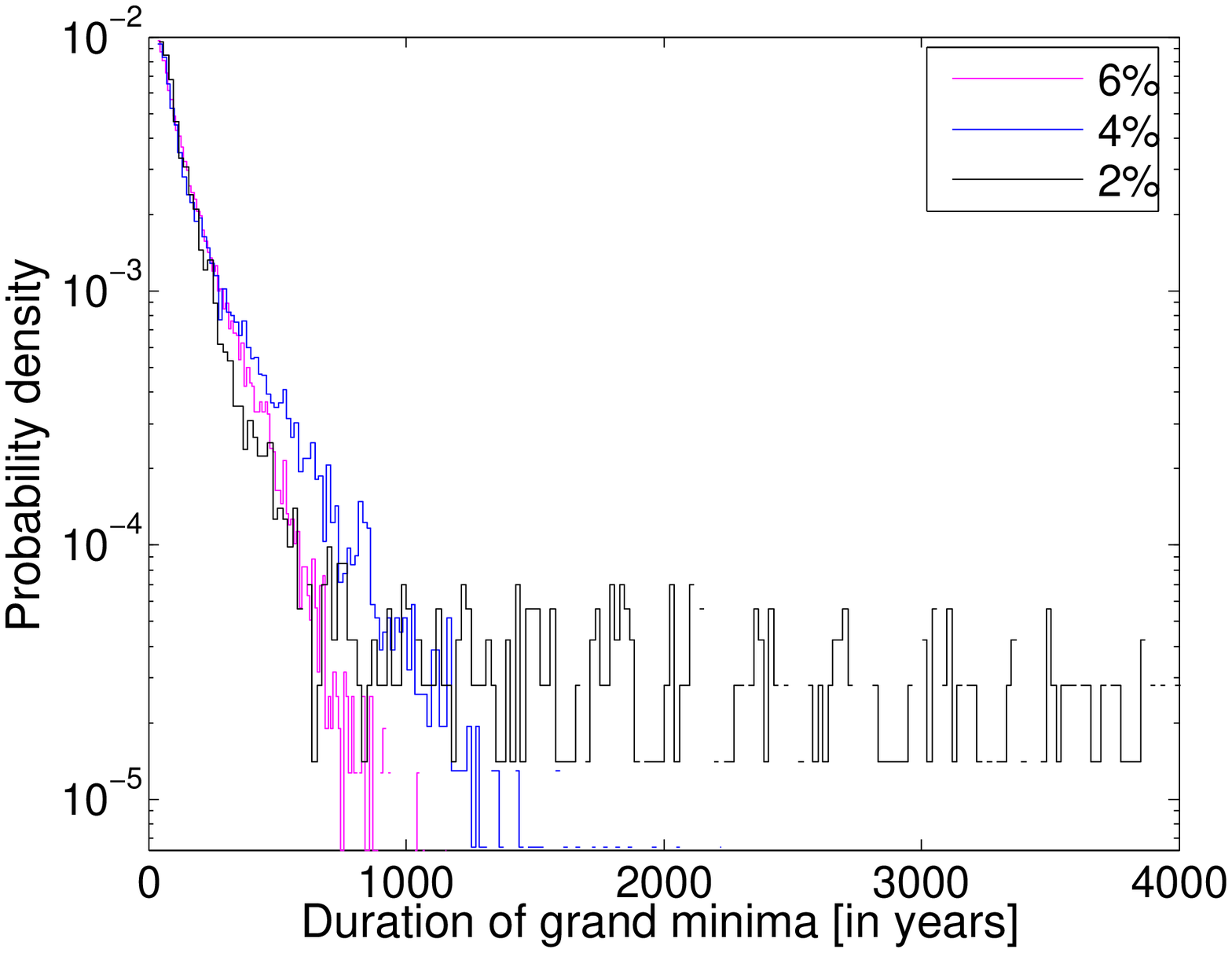}\includegraphics[width=0.48\textwidth]{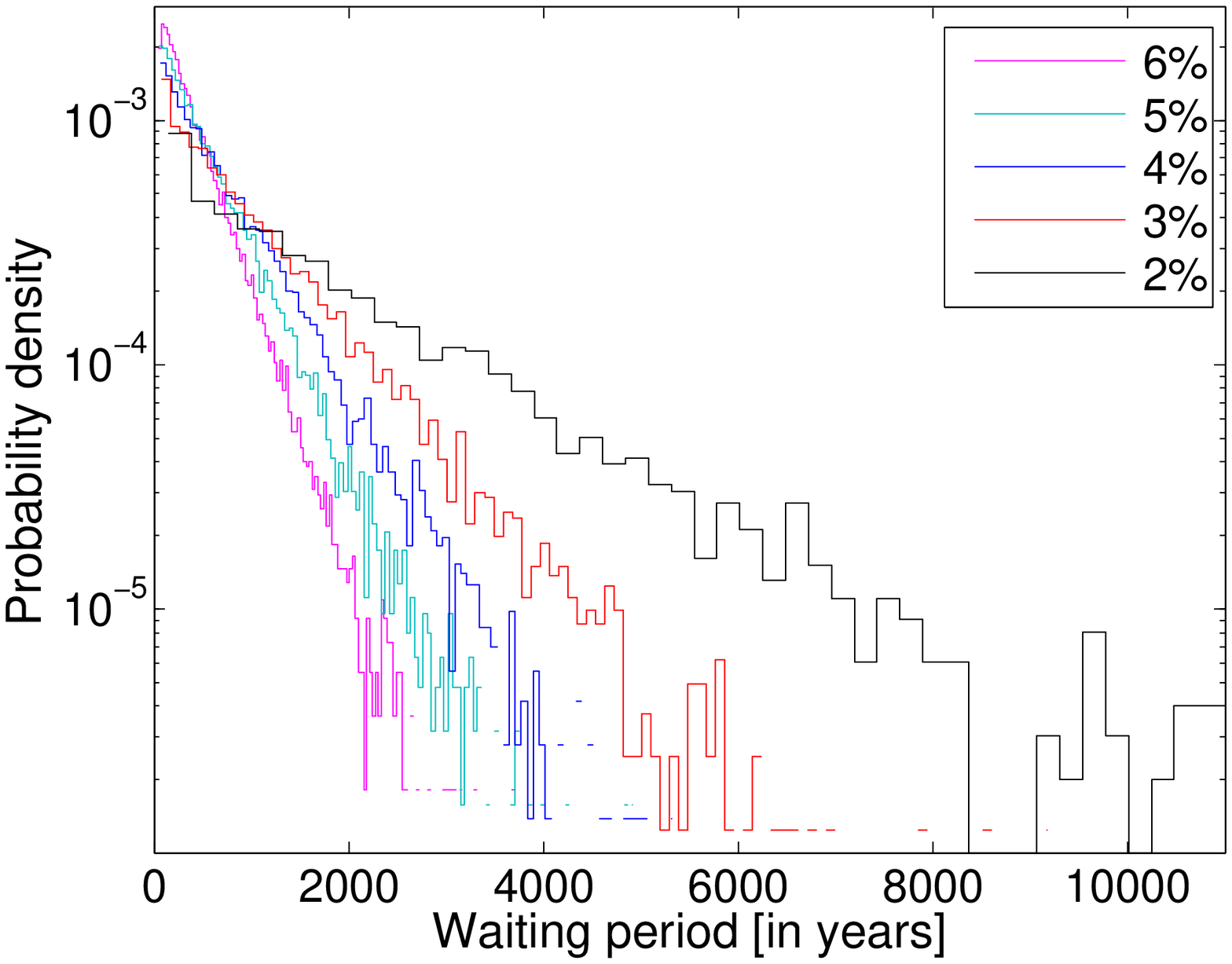}
\caption{\label{MagNoiseDistribitions}Probability density of duration of grand minima (on the left) and of waiting period for grand minima ((on the right) when the stochastic magnetic noise of different amplitudes (presented in the legend) are incorporated in the time delay BL dynamo model. The simulations are run for 10 million years to obtain a statistical distribution of the duration of grand minima and its waiting period. Both of these periods shorten on increasing the level of noise from 2\% to 6\%, implying that the noise recovers the cycle from grand minima as well as it triggers such episodes. 30\% fluctuations are used in $\alpha_{BL}$ with $\alpha_{mean}=0.174$ (${N_D}^{mean}=13.3$) for these simulations.}
\end{figure*}
Having just overcome the hurdles of recovering the cycle from grand minima and mimicking its occurrence statistics by the noise-assimilated time delay BL dynamo model, we pose some challenges before it. Can the model exhibit bimodal distribution of the probability density of decadal sunspot number, which was recently found in the reconstructed solar activity record by \citet{usoskin_evidence_2014}? Is the model capable of reproducing the collapse in phase space (an experimental signature of the Maunder minimum), which was noticed by \citet{passos2011grand} by building the phase space using the sunspot number as the proxy of the toroidal field? Interestingly, we find that the incorporation of magnetic noise in the model to explain grand minima seems to be vital, not only because it is theoretically plausible but also because it generates magnetic cycles, which show a bimodal distribution in probability density of magnetic energy proxy (see Figure \ref{Bimodal_distribution}). This implies that the grand minima mode is not a mode of solar activity that results due to some random fluctuations in the regular activity mode, but it is itself a separate mode of dynamo operation. In other words, the grand minima cannot be considered as events that correspond to a tail of a single regular mode of solar activity. This bimodal nature in the probability density of smoothed SSN is actually due to the hysteresis present in the model \citep{kitchatinov_parametric_2015}. It should be noted that the hysteresis is a phenomenon where bistable solutions exist in a certain parameter regime and either of them is realized depending upon the system's initial conditions. And these bistable solutions emanate
in our model due to the consideration of the lower operating threshold in the Babcock-Leighton dynamo. If we do not consider this lower operating threshold, there exist no decaying solutions and no hysteresis, which consequently blows out the bimodal structure of probability density of smoothed SSN. So, we speculate that it might be this lower operating threshold which had left its imprint on the reconstructed SSN and now this threshold is being revealed in the bimodal distribution of the reconstructed SSN. Moreover, it also qualitatively reproduces the phase space collapse (see Figure \ref{Phase space collapse}), which \citet{passos2011grand} tried to achieve using their low order dimensional model but failed to do so by using the stochastic fluctuations in their model's linear $\alpha$-effect. The collapse in the phase space signifies that the Maunder-minimum episode resulted as a strong intermittency in magnetic cycles. It is interesting to note in Figure \ref{Phase space collapse} (on the right of the upper panel) that the polarity reversal of magnetic cycles still happen during grand minimum even when Babcock-Leighton source switches off. It is because the stochastically injected near-surface poloidal fields due to 'magnetic noise' are still being carried down to the base of the convection zone where these weak fields are sheared due to the differential rotation to regenerate toroidal fields.\\

We further explore the effect of the level of noise on the duration of quiescent phases (grand minima) and their frequency (Figure \ref{MagNoiseDistribitions}, left). On increasing the level of noise from 2\% to 6\% (of the average of amplitude of the toroidal field during regular activity cycles), it is seen that the average length of the quiescent phases shortens. This is reasonable because as the level of noise is increased, the cycle recovers sooner from a grand minimum. We also observe that there exists a threshold on the level of noise below which the cycle does not recover in a finite time. However, we do not attempt to find that particular threshold. Similarly, we study the effect of the level of noise on the waiting period (Figure \ref{MagNoiseDistribitions}, right) of grand minima and observe that the increasing level of noise again shortens the average waiting period. (Waiting period for grand minima is the time interval between the occurrence of two successive grand minima.) So, the noise too has a handle over the waiting period for grand minima. It immediately infers that the noise can also take part in inducing grand minima, which strikingly contrasts to the results reported by \citet{charbonneau_multiperiodicity_2001} using a one-dimensional iterative map. Since noise too can trigger grand minima, predicting such events can be more difficult than hitherto thought.\\


\section{Conclusions} \label{section conclusions}
To sum up, we have probed the stochastically forced time delay solar dynamo model using numerical bifurcation analysis and found that the time delay Babcock-Leighton model exhibits hysteresis phenomenon, which can be directly related to the two distinct modes of dynamo operation in the Sun \citep{usoskin_evidence_2014}. Such hysteresis behavior was shown to exist in complicated mean-field dynamo models with the transport coefficients depending upon the strength of the magnetic field \citep{kitchatinov_dynamo_2010} and even in direct numerical simulations of 3D turbulent dynamos \citep{karak_hysteresis_2015}. However, the model suffers from a problem of unsuccessful recovery once the Sun is pushed into a grand minimum. We found that the consideration of a low-amplitude stochastic 'magnetic noise' as an additional poloidal source in the Babcock-Leighton model recovers the solar cycle every time the cycle is kicked into a grand minimum and reasonably mimics the occurrence statistics of the grand minima inferred from the cosmogenic radionuclides. It also exhibits bimodal distribution of toroidal magnetic energy (a proxy of sunspot number) implying the presence of two distinct modes of activity \citep{kitchatinov_parametric_2015} and reproduces the phase space collapse, which is a signature of the only directly observed grand minimum, i.e. the Maunder minimum. We observed that the duration of grand minima is determined by the level of noise and the higher level of noise pulls the Sun back to regular activity sooner. Interestingly, we found that the noise also has a handle over the waiting period for grand minima, which strikingly contrasts to the results suggested by \citet{charbonneau_multiperiodicity_2001} using a one-dimensional iterative map. Our result implies that the noise due to small-scale turbulent dynamo action or other sources can sometimes shut down the global dynamo, triggering grand minima. We also saw that the solar cycle shows intermittent behavior, wandering between two dynamically distinct states.\\

All the results presented in this paper were obtained by using a coherence time of 4 years for the fluctuations in both the poloidal sources- the Babcock-Leighton source and the additive 'magnetic noise'. We note that the shorter coherence time (0.1 years) for the fluctuations in magnetic noise if used in our model, exhibits grand minima of very long duration when the cycle's amplitude falls below the sunspot formation threshold. Such result is simple to understand as the consideration of shorter coherence time implies that the random noises that are invoked after each coherence time cancel each other's effect and hence the cycle has to wait a long time to come out of such grand minima. We performed several simulations with increasing coherence time (from 0.1 to 4 years) for fluctuations in magnetic noise and found that the average duration of grand minima shortens from thousands of years (for a coherence time of 0.1 years) to a couple of hundreds of years (for a coherence time of 4 years). It should be noted that the duration of grand minima also shortens on increasing the level of noise imposed.\\ 

We demonstrated that the sun can be pushed into grand minima epochs by the the fluctuations in tilt angles of bipolar active regions, a change in flux transport timescale, and/or magnetic noise. However, the recovery from such reduced activity phases is richly facilitated by the noise. Therefore, we conclude that the magnetic noise due to small-scale turbulent dynamo action in the solar convective zone or due to other sources plays a vital role in the origin and recovery from grand minima and is important to understand the grand minima in the sun and the solar-type stars. 

\acknowledgments
We acknowledge support from the Center of Excellence in Space Sciences India (CESSI) and the Ministry of Human Resource Development, India. B. T. wishes to thank IISER Kolkata for providing Summer Research Internship and the CESSI's research group for illuminating discussions. 

\software{MATLAB and dde23 package}



\end{document}